\documentstyle[prd,aps,amstex,preprint]{revtex}
\tightenlines

\def\A{{\cal A}}
\def\L{{\cal L}}

\begin{document}


\preprint{YITP-00-45}

\title{Gravitational waves in cosmological models of \\
Ho\v{r}ava-Witten theory}\author{Osamu Seto and Hideo Kodama}
\address{Yukawa Institute for Theoretical Physics, Kyoto University\\
Kyoto 606-8502, Japan}
\date{\today}

\maketitle


\begin{abstract} 

We study the behavior of gravitational waves and their backreaction on the 
background in cosmological solutions of the five-dimensional 
Ho\v{r}ava-Witten theory. As a dynamical background, we consider two 
cosmological solutions with spatially flat expanding FRW branes, called 
$(\uparrow)$- and $(\downarrow)$-solutions, in which the orbifold size 
increases and decreases in time, respectively. 
For these background solutions, the wave equation for the tensor 
perturbation can be solved by the method of separation of variables, and 
the mode functions are classified by a separation constant which can be 
regarded as a graviton mass.  We show that the spatial behavior of the 
mode functions are the same for both background solutions, but the 
temporal behavior is significantly different. We further show that for the 
$(\uparrow)$ solution, the background bulk geometry is unstable against 
the backreaction of the perturbation, while for the $(\downarrow)$ 
solution, the backreaction on the bulk geometry can be neglected. We also 
show that, in contrast to the effect to the bulk geometry, the 
backreaction of the perturbation significantly alters the intrinsic 
geometry of the brane for the $(\downarrow)$ solution.
\end{abstract}

\pacs{PACS numbers: }


\section{Introduction}

Over the past few years, a considerable number of studies have been made 
on the brane-world scenario in which our universe is realized as a 
boundary of a higher dimensional spacetime 
\cite{brane,hw,witten,lukas1,lukas2,reall,ell,lukas3,lidsey,dabrowski,meissner,RS1,RS2,gt-etc,gs-etc,cline-etc}.
In particular, inspired by the recent progress in the heterotic M-theory
\cite{hw}, five-dimensional brane-world models in which 3-branes are 
embedded in an effective five-dimensional spacetime compactified 
on $S^1/{\Bbb Z}_2$ \cite{witten,lukas1} have attracted much attention. 
For example, cosmological solutions in the five-dimensional 
Ho\v{r}ava-Witten theory have been 
discovered\cite{lukas2,reall,ell,lukas3,lidsey}. The five-dimensional 
models of Randall and Sundrum, which were proposed to solve the Hierarchy 
problem\cite{RS1} and to demonstrate an alternative to compactification 
\cite{RS2}, also have many similarities to the five-dimensional 
Ho\v{r}ava-Witten theory. %

In the brane-world scenario, all ordinary matter fields are confined
on the brane, while graviton can propagate in the fifth-dimension. 
Hence, in order to test the idea of the brane-world, 
one needs to study the nature of the gravity in this scenario. 
The behavior of linearized gravity in the Randall - Sundrum models 
has recently been studied in detail. It has been shown that massless 
modes of the metric perturbation are decoupled from  massive modes and 
the Einstein gravity is recovered at low energy scales (See, 
\textit{e.g.}\cite{RS1,RS2,gt-etc,gs-etc}). However these
investigations have been
done only in the highly symmetric background models such that
four-dimensional maximally symmetric branes, \textit{i.e.}, Minkowski
brane \cite{RS1,RS2} and de Sitter brane \cite{gs-etc}, are embedded
in a five-dimensional anti-de Sitter spacetime, which also is locally
maximally symmetric in five-dimensions.       

On the other hand, many people have discussed the possibility of a
homogeneous and isotropic Friedmann-Robertson-Walker (FRW) brane-world
in the Ho\v{r}ava-Witten theory \cite{lukas2,reall,ell} and in Randall
and Sundrum's scenario \cite{cline-etc}. However, there have so far
been few attempts to investigate perturbations on such a dynamical
cosmological brane-world, although the four-dimensional real universe has
inhomogeneous fluctuations as is shown by the observations of the CMB
anisotropy. 

As pointed out in Ref. \cite{kodama}, it is expected that all
gravitons on the brane become massive in a dynamical brane-world
model. Then, the excitation of massive graviton would become a crucial
defect in the brane-world scenario, or provide a new model of dark
matter in the brane-world cosmology. It is therefore important to
study the evolution of perturbations on a dynamical brane model in
order to explore the cosmological consequences of the brane-world
idea. 

Recently, formalisms for cosmological perturbations on 
brane-world have been developed by several authors
\cite{kodama,mukoh,Maa,Lan,Bra,koyama}. In particular, for general 
cosmological brane-world models, the evolution equations for the metric 
and matter perturbations in the bulk and the boundary conditions for them 
at the brane have been established by Kodama \textit{et al}.\cite{kodama} 
in terms of gauge-invariant variables. 
The perturbations on a brane are inevitably coupled to the perturbations 
on the bulk. The evolution equations and boundary conditions for
cosmological perturbations, in particular, the scalar and vector
perturbations on the cosmological brane are too complicated to solve. 
On the other hand, as shown in \cite{kodama}, when the anisotropic
stress perturbation vanishes, the tensor perturbations decouple from
the matter perturbation and the boundary condition becomes a Neumann
type. Hence, as far as the tensor perturbations are concerned, the
problem is easier to deal with.  

In this paper, as a first step to investigate the perturbations 
on a dynamical brane-world, we study the behavior of 
tensor perturbations on two cosmological brane-world models  
in five-dimensional Ho\v{r}ava-Witten theory.  
In general, the equations for perturbations are no longer separable
for a dynamical 3-brane which is not maximally symmetric as a
hypersurface. Fortunately, for the cosmological solutions in the 
Ho\v{r}ava-Witten
model found in \cite{lukas2,reall}, 
the evolution equation of tensor perturbations becomes separable. However 
the evolution equation is still rather complicated to solve exactly. So, 
in this paper, we consider the late time behavior of the tensor 
perturbation by using WKB approximation to analyze the evolution equation. 
Further we discuss the backreaction problem and the stability of the 
background solutions as well as the brane motion using the second-order 
perturbation theory. We shall show that one of the cosmological solution 
is unstable if the backreaction of the tensor perturbation is taken into 
account. We shall also show that the backreaction from massive modes 
significantly alters the brane motion, i.e., the evolution of the brane 
universe, for the solution whose bulk geometry is stable against the tensor perturbation.

The present paper is organized as follows. 
In the next section we briefly recapitulate the five-dimensional
Ho\v{r}ava-Witten theory and the cosmological solutions which 
are used as the background for perturbation. In Sec. III we give 
the equations of motion for the tensor perturbation and the boundary 
condition for them, and analyze the behavior of the tensor 
perturbation by using WKB method. Then in Sec. IV we discuss the 
backreaction of the perturbation on the background. Section V is devote to 
conclusion and discussion.


\section{Cosmological solutions of Ho\v{r}ava-Witten theory} 

Ho\v{r}ava and Witten have shown that the strongly coupled limit of
the $E_8\times E_8$ heterotic string theory has been identified with
M-theory compactified on a $S^1/{\Bbb Z}_2$ orbifold with $E_8$ gauge
fields on each orbifold fixed plane \cite{hw}. After a
compactification on a Calabi-Yau three-fold, the fields of the
standard model can be confined to 3-brane
\cite{witten}. Thus one has an effective model in which our
four-dimensional universe is a 3-brane embedded in an effective
five-dimensional spacetime compactified on $S^1/{\Bbb Z}_2$. As was shown by Lukas, Ovrut, Stelle and Waldram\cite{lukas1}, the bosonic sector of this effective model contains the 5-dimensional metric $g_{MN}$, a modulus $V$ describing the variation of the Calabi-Yau volume along the orbifold, and the $U(1)$ gauge field $\A_M$ and two charged scalars $(\sigma, \xi)$ parametrizing the anti-symmetric tensor field.  If we assume that the gauge field and the two charged scalar fields vanish, the five-dimensional effective action for the bosonic sector of the  Ho\v{r}ava-Witten theory is given by \cite{lukas1}:
\begin{eqnarray}
S &=& \frac{1}{2 \kappa_5^2} \left
[ \int_{M_5}\sqrt{-g}\left(R-\frac{1}{2V^2}(\nabla V)^2
-\frac{\alpha^2}{3V^2}\right) \right. \nonumber \\ 
 & & \left. +2\sqrt{2}\int_{M_4^{(1)}}\sqrt{-g}\frac{\alpha}{V}
 -2\sqrt{2}\int_{M_4^{(2)}}\sqrt{-g}\frac{\alpha}{V} 
 \right] , \label{action}
\end{eqnarray}
where $\kappa_5$ is the five-dimensional gravitational constant, 
$\alpha$ is a constant, and ${\mathcal M}_5$ is the five dimensional 
spacetime bounded by the branes ${\mathcal M}_4^{(1)}$ and 
${\mathcal M}_4^{(2)}$. 

The cosmological solutions with flat FRW branes for this effective action have been constructed by Lukas \textit{et.al.} \cite{lukas2}
and been generalized to include closed and open FRW branes by 
Reall \cite{reall}. In the present paper, as a simple case, 
we shall consider only the model in which 
expanding flat FRW branes are embedded in the five-dimensional bulk. 

Let $x^4 \equiv y$ be a coordinate in the orbifold direction with
$y\in[-\pi\rho,\pi\rho]$ and ${\Bbb Z}_2$ acting on $S^1$ by
$y\rightarrow-y$. The orbifold fixed planes are located at
$y=0,\pi\rho$. Then, starting from the effective five-dimensional action
(\ref{action}) with the metric ansatz: 
\begin{equation}
ds^2 = a_0^2e^{2A(t)} C(y)(-dt^2 + \delta_{ij}dx^idx^j) 
       + e^{2B(t)} D(y) dy^2 \,  , 
\label{metric}
\end{equation}
one obtains the cosmological solutions of five-dimensional Ho\v{r}ava-Witten theory with flat FRW branes \cite{lukas2}: 
\begin{equation}
C(y) = D(y)^{1/4} = \frac{\sqrt{2}}{3} \alpha |y| + 1 \,, 
\label{ymetric}
\end{equation}
\begin{equation}
e^{2A} = t^{1-\delta}, \quad e^{2B} = t^{2\delta}, 
\end{equation}
where $a_0$ are constants and $\delta = \pm \sqrt{3}/2$. The field $V$ for these solutions is given by 
\begin{equation}
V=e^{B(t)}C(y)^3.
\end{equation}
Hereafter we shall refer to the upper and lower choices of sign as 
the $(\uparrow)$ and $(\downarrow)$ solutions, respectively. 

The $(\downarrow)$ solution describes the model that 
the four-dimensional FRW universe expands while the orbifold space shrinks. 
On the other hand, the $(\uparrow)$ solution describes the model 
in which both the four-dimensional FRW universe and the orbifold space 
expand, and the latter expands faster than the former does. 

For a while, we shall focus our attention on the four-dimensional
brane at $y=0$.  In terms of the cosmic proper time $\tau$ defined by 
\begin{equation}
\tau \equiv a_0 \int e^A dt = \frac{2a_0}{3-\delta} t^{(3-\delta)/2} . 
\end{equation} 
The Hubble parameter is given by 
\begin{equation}  
  H(t) \equiv { 1\over a}{da \over d\tau} 
        = \frac{1-\delta}{2a_0} t^{-(3-\delta)/2} \,. 
\end{equation}  
Then the wave-number $k_H$ whose wavelength corresponds to the horizon 
scale is given by 
\begin{equation}  
  k_H \equiv a H = \frac{1-\delta}{2t} \,. 
\end{equation}  
The scale factor of the four-dimensional FRW brane is written as 
\begin{equation}  
 a(\tau) = a_0 \left( 
                      \frac{3-\delta}{2a_0}\tau 
               \right)^{ (1-\delta)/(3-\delta)} \,. 
\end{equation}  
Comparing this scale factor with that of the no-extra-dimension 
cosmology, we find that the brane expands as if it were a standard
four-dimensional flat FRW universe which contains a perfect fluid obeying 
the equation of state $p = w \rho $ with 
\begin{equation}  
  w = \frac{3+ \delta}{3(1-\delta)} \,, 
\end{equation}  
although the brane-world solutions considered here are vacuum solutions. 
In this picture, the FRW brane in the
$(\uparrow)$ solution looks like an unphysical universe because the
dominant energy condition violates, $ w > 1$ . On the other hand, in
the $(\downarrow)$ case, the FRW brane describes a physical universe
in the sense that the dominant energy condition holds, $ 0 < w \approx
0.37 < 1$.


\section{Behavior of tensor perturbations}

\subsection{Wave equation for tensor perturbations}

Gravitational perturbations in the bulk are decomposed into components of 
the three types, scalar, vector and tensor, and each component can be
expanded by tensor harmonics of the same type on the 3-space of constant
curvature. Then, the tensor perturbations represent gravitational wave
modes in the four-dimensional FRW brane. 

For the action (\ref{action}), the anisotropic stress perturbation vanish 
both in the bulk and in the brane. Then, for the background metric
Eq.(\ref{metric}) and (\ref{ymetric}), the equation of motion for
the tensor perturbation and the boundary condition are given,
respectively, by \cite{kodama}
\begin{equation}
\ddot{H}_T+\left(2\dot{A}(t)+\dot{B}(t)\right)\dot{H}_T-\frac{a_0^2
e^{2(A(t)-B(t))}}{C(y)^3}H''_T+ k^2 H_T=0,
\label{eomH}
\end{equation}
\begin{equation}
H_T{}' = 0 , \quad at \,\, y = 0, \pi\rho,
\label{bc}
\end{equation}
where $H_T$ is the expansion coefficient of the metric 
perturbation,$\delta g_{ij} = 2 a_0^2 e^{2 A} C(y) H_T 
\mathbb{T}$${}_{ij}$,
in terms of the tensor-type harmonic tensor $\mathbb{T}$${}_{ij}$ on
the flat 3-space \cite{kodama}. Here, $-k^2$ is an eigenvalue of the
d'Alembertian on the flat 3-space, and dots and primes denote
derivatives with respect to $t$ and $y$, respectively. 

Note that the boundary condition (\ref{bc}) is simply written as the 
derivative with respect to $y$ as in the static brane case. 
However, the evolution equation (\ref{eomH}) contains an additional 
friction term $\dot{B} \dot{H}_T$, which does not exist when the 
background orbifold space is static, and the $y$-derivative term (the 
third term), which gives graviton's mass as we shall see below. 
So, these two terms reflect the effect of the dynamics of the background 
branes model on  gravitational waves. 

Provided that $H_T(t,y)=T(t)Y(y)$, the equations of motion
(\ref{eomH}) are reduced to the following set of equations for $Y(y)$ and 
$T(t)$:
\begin{equation}
Y''_l+\left(\frac{m}{a_0}\right)^2 C(y)^3 Y_l = 0, 
\label{eomY}
\end{equation}
\begin{equation}
\ddot{T}_l+\left(2\dot{A}(t)+\dot{B}(t)\right)\dot{T}_l+ k^2 T_l+m^2 e^{2(A(t)-B(t))} T_l=0, \label{eomT}
\end{equation}
where the dimensionless constant $m$ is defined as follows:
\begin{equation}
m^2 \equiv \left(\frac{\sqrt{2}\alpha l a_0}{3}\right)^2 .
\label{m}
\end{equation}
Here, $l$ represents the level of inhomogeneity in the orbifold
direction, and $m$ takes discrete values labeled by an integer $n$, as 
we shall see in the next subsection. Note that the signature of the
second term in LHS of Eq.(\ref{eomY}) is chosen so that the solutions
of Eq.(\ref{eomY}) satisfy the boundary condition (\ref{bc}). 

In the case of $\dot{B} = 0$, the last term in Eq.(\ref{eomT})
provides the eigenvalue (times $e^{2A}/a_0^2$) of the d'Alembertian on
the four-dimensional brane. Therefore the $m = 0$ modes behave as the
massless mode in the brane when the orbifold space is static. In this
sense, we shall refer to $m = 0$ $(l = 0)$ and $m\neq 0$ $(l \neq 0)$
modes as ``massless'' and ``massive'' modes, respectively.

\subsection{Solutions of $y$-dependent part}

In this subsection, we shall give the solutions of Eq.(\ref{eomY})
which satisfy the boundary condition (\ref{bc}). 

For $m=0$ modes, Eq.(\ref{eomY}) reads 
\begin{equation}
Y_l(y) = C_1 + C_2\, y , 
\end{equation}
where $C_1$ and $C_2$ are integration constants. From the boundary
condition (\ref{bc}), we find that $C_2$ must
vanish. Therefore, the zero-mode($l=0$) solution for $y$-dependent part is
\begin{equation}
Y_l(y) = C_1 . 
\end{equation}
On the other hand, for $m \neq 0$ modes, the solutions of Eq.(\ref{eomY}) for $y \geq 0$
are given by
\begin{equation}
Y_l(y) = C(y)^{1/2}\left[C_1 H^{(1)}_{1/5}\left(\frac{2l}{5}C(y)^{5/2}\right)+C_2 H^{(2)}_{1/5}\left(\frac{2l}{5}C(y)^{5/2}\right)\right],
\end{equation}
where $C_1$ and $C_2$ are constants. 

From the boundary condition at $ y=0 $, we find that the ratio of $C_1$
and $C_2$ becomes
\begin{equation}
\frac{C_2}{C_1} = -\left.\frac{H^{(1)}_{-4/5}(z)}{H^{(2)}_{-4/5}(z)}\right|_{z=2l/5} 
\end{equation}
for each $l$. From the boundary condition at $y=\pi\rho$, we obtain
\begin{equation}
\left.\frac{H^{(1)}_{-4/5}(z)}{H^{(2)}_{-4/5}(z)}
\right|_{z=2l/5} =
\left. \frac{H^{(1)}_{-4/5}(z)}{H^{(2)}_{-4/5}(z)}
\right|_{z=2l(\sqrt{2}\alpha\pi\rho/3+1)^{5/2}/5} .
\end{equation}
This gives the value of $l$ for each excited mode.

For the case of $2l/5 \gg 1$ , $l$ can be written as 
\begin{equation}
l \approx \frac{5}{2}\frac{n\,
\pi}{\left(\sqrt{2}\alpha\pi\rho/3+1\right)^{5/2}-1},
\label{l}
\end{equation}
where $ n = 1,2,\cdots$. $\alpha\pi\rho \ll 1$ is satisfied in the
context of five-dimensional Ho\v{r}ava-Witten theory, except for
inflationary epoch \cite{lukas3}. In this case, Eq.(\ref{l}) 
reads 
\begin{equation}
m \approx \frac{a_0 \, n }{\rho}.
\label{l1}
\end{equation}
Thus the reduced KK ``mass'' defined by
\begin{equation}
m_{KK,l} \equiv  \frac{m}{a_0 e^{B(t)}\sqrt{C(y)}},
\end{equation}
takes a typical value of the orbifold energy scale.

The norm squared of $Y_l(y)$ is then given by
\begin{equation}
|Y_l(y)|^2 \approx \cos^2 \left[\frac{2l}{5}\left\{\left(\sqrt{2}\alpha 
|y|/3+1\right)^{5/2}-1\right\}\right] .
\end{equation}
Therefore $|Y_l(y)|^2$ has peak at each boundary (at $ y=0 $ and $ y 
= \pi\rho $).  

\subsection{Solutions of $t$-dependent parts}

In this subsection, we shall examine the behavior of solutions to 
Eq.(\ref{eomT}). For simplicity, we omit the suffix $l$ hereafter. 

For the $m=0$ modes,  Eq.(\ref{eomT}) is exactly solved to yield%
\begin{equation}
T=D_1 H_0^{(1)}(kt)+D_2 H_0^{(2)}(kt),
\label{sol}
\end{equation}
where $D_1$ and $D_2$ are constants. 
In contrast, for the $m \neq 0$ modes, Eq.(\ref{eomT}) cannot be solved 
exactly. Therefore, we analyze the behavior of solutions by means of the 
WKB method by rewriting Eq.(\ref{eomT}) as 
\begin{equation}
\left(t^{1/2}T\right)\ddot{}+S(t)^2 t^{1/2}T = 0 , 
\label{T:1}\end{equation}
where $S(t)$ is defined by
\begin{equation}
S(t) \equiv \sqrt{k^2+\frac{1}{4 t^2}+m^2 t^{1-3\delta}}.
\label{S}
\end{equation}
In the region where $ |\dot{S}| \ll S^2 $ holds, we can use the WKB method to  
obtain the approximate solution
\begin{equation}
T(t) \approx D_1(k) (t S(t))^{-1/2} \exp\left[-i \int^t S(t') dt' 
\right]+D_2(k) (t S(t))^{-1/2} \exp\left[i \int^t S(t') dt' \right].
\label{WKB}
\end{equation}
Since $\dot S/S^2$ is written as
\begin{equation}
\frac{|\dot{S}|}{S^2} = \frac{2\left|
-1+2(1-3\delta) m^2 t^{3(1-\delta)} \right|}{\left(1+4(kt)^2+4 m^2
t^{3(1-\delta)}\right)^{3/2}} ,
\end{equation}
the WKB approximation is good in the region where $t \gg
m^{-2/(3(1-\delta))}$ or $t \gg 1/(2k)$. 
The former relation, which is equivalent to $m_{\rm KK}\tau\gg1$, is 
satisfied when the time scale is larger than the orbifold radius, while 
the latter is satisfied when the wavelength is shorter than the Horizon 
radius, \textit{i.e.}, $k \gg k_H$. 

In the $(\uparrow)$ background case, since $t^{1-3\delta}$ is a
monotonically decreasing function, the mass term becomes negligible
compared with the $k^2$-term on the right hand of
Eq.(\ref{S}). Therefore, solutions to Eq.(\ref{T:1}) are well
approximated by the solutions Eq.(\ref{sol}) in the massless case in a
sufficient late time for any fixed $k$. In contrast, in the
$(\downarrow)$ background case, $m^2t^{1-3\delta}$ term increase with
time, and the solutions deviate from those for the massless case in
late times. In particular, for $4m^2t^{3(1-\delta)}\gg1+4(kt)^2$, the
WKB solution is given by
\begin{equation}
T \approx (m t^{3(1-\delta)/2})^{-1/2} \left( 
D_1(k) \exp\left[-i\frac{2m t^{3(1-\delta)/2}}{3(1-\delta)}\right]+D_2(k)  
\exp\left[i \frac{2m t^{3(1-\delta)/2}}{3(1-\delta)}\right] \right),
\label{late}
\end{equation}
after an appropriate redefinition of the constants $D_1$ and $D_2$.
Note that the late time solution (\ref{late}) does not depend on the 
wavenumber $k$ explicitly, and the argument of the exponential function in 
(\ref{late}) is proportional to $m_{KK}\tau $.
In particular, from Eq.(\ref{sol}) and Eq.(\ref{late}) or 
from Eq.(\ref{S}) and Eq.(\ref{WKB}), we see that the ratio of the 
amplitude of a massive mode to that of a massless mode behaves as
\begin{equation}
\frac{T_{m \neq 0}}{T_{m = 0}} \propto S(t)^{-1/2} \sim e^{(B(t)-A(t))/2} 
=t^{(3\delta-1)/4}
\end{equation}
for $m_{KK} \gg k/a$. This apparently shows that the massive modes become 
negligible in late times. However, if we consider their backreaction, the 
conclusion changes significantly, as we will see in the next section. 

\section{Backreaction of the perturbation}

In this section, we study the back reaction of the tensor perturbation on 
the bulk background geometry and on the intrinsic geometry of the brane 
with the help of the second-order perturbation theory. 

First note that if we expand the deviation of the bulk geometry from the 
background in terms of some small parameter, the second-order part 
$\delta_2 g$ satisfies the equation
\begin{equation}
(\L^{(1)}\delta_2 g)_{MN}=\kappa_5^2 (T^{\rm GW}{}_{MN}
+\delta_2 T_{MN}),
\end{equation}
where $\L^{(1)}$ is the differential operator for the metric
perturbation obtained from the linear perturbation of the Einstein
equations, $T^{\rm GW}{}_{MN}$ is the effective energy-momentum tensor
for the linear perturbation $\delta_1 g$ of geometry, which is
quadratic in $\delta_1 g$, and $\delta_2 T_{MN}$ is the second-order
perturbation with respect to the field $V$ of the bulk energy-momentum
tensor
\begin{equation}
\kappa_5^2 T_{MN}=\frac{1}{2V^2}\partial_M V \partial_N V
-\frac{1}{2}g_{MN}\left(\frac{1}{2V^2}(\nabla V)^2
+\frac{\alpha^2}{3V^2}\right).
\end{equation}
The explicit expression for $T^{\rm GW}{}_{MN}$ is given in the Appendix. 
In contrast to the linear perturbation, the spatial average in the 
3-dimensional sense does not vanish in general and produces a spatially 
homogeneous contribution to $\delta_2g$. This contribution can be regarded 
as the backreaction of the perturbation on the background geometry. 

In particular, for the tensor perturbation in the models considered 
in the present paper, the effective energy density $\rho_{\rm GW}$ is 
given by
\begin{eqnarray}
2\kappa_5^2\rho_{\rm GW} &\equiv& -2\kappa_5^2\langle T^{\rm GW}{}^0_0 
\rangle \\
 & = & (a_0^2 e^{2A(t)} C(y))^{-1}
 ( |\dot{H_T}|^2 + k^2|H_T|^2 ) + e^{-2B} |H'_T|^2 +\cdots 
\label{ener}, 
\end{eqnarray}
under the normalization of the tensor harmonics as $\langle {\mathbb 
T}_{ij}{\mathbb T}^{ij}\rangle =1$. The leading term of $\rho_{\rm GW}$ 
is given by the first term, which behaves as $e^{-2A(t)}\dot{H_T}^2 \sim 
t^{-2+\delta} S$. In the meanwhile, the leading term for the energy density 
of the $V$ field determining the bulk background geometry is given by the 
potential energy $\alpha^2/(3V^2) \propto t^{-2\delta}$, and its  
second-order perturbation is given by
\begin{equation}
-\kappa_5^2 \delta_2 T^0_0=2\frac{\delta_2 V}{V}\kappa_5^2T^0_0
+\frac{\dot B^2}{2a^2}\frac{(\delta_2\dot V)}{\dot V}
+\frac{\alpha^2}{V^2}\frac{\delta_2 V'}{V'}
\sim 2\frac{\delta_2 V}{V}\kappa_5^2T^0_0.
\end{equation}
Comparing these energy densities, we find that the effect of $\rho_{\rm 
GW}$ becomes negligible in the $(\downarrow)$ background case if 
$\delta_2 V/V$ is small. Since the field equation for $V$ is given by
\begin{equation}
\square_{g+\langle\delta_2 g\rangle}\left(\ln(V+\langle\delta_2 
V\rangle)\right)
+\frac{2\alpha^2}{3(V+\langle\delta_2V\rangle)^2}=0,
\end{equation}
up to the second order, $\langle\delta_2 V\rangle /V$ is small if 
$\langle\delta_2 g\rangle$ is negligible. Hence, in this case the
backreaction can be neglected in late times. In contrast, for the
$(\uparrow)$-model, the decrease of $\rho_{\rm GW}$  is slower than
$-T^0_0$. Hence this model is unstable against the backreaction of the
tensor perturbation. 

Next, we consider the backreaction effect on the intrinsic geometry of the 
brane. Since a full treatment of this problem is very difficult, we only 
make rough estimate using the Hamiltonian constraint along the 
brane,
\begin{equation}
{}^{(4)}\!R = -2\kappa_5^2T_{\perp\perp}+ K^2- K^{\mu\nu} K_{\mu\nu},
\end{equation}
where ${}^{(4)}\!R$ is the Ricci scalar of the four-dimensional metric 
$g_{\mu\nu}$ of the brane, $T_{\perp\perp}$ is the component of the 
energy-momentum tensor along the unit normal to the brane, and $K_{\mu\nu}$ 
is the extrinsic curvature of the brane. As explained above, if we take 
into account the backreaction of the tensor perturbation on the brane 
geometry, $T_{\perp\perp}$ should be replaced by $T^5_5+T^{\rm GW}{}^5_5 
+\delta_2 (T_{\perp\perp})$ in the second-order perturbation framework. 
Here, $K^\mu_\nu$ is related to the intrinsic energy-momentum tensor 
${}^{(4)}\!T^\mu_\nu$ of the brane by the junction condition 
\begin{equation}
K^\mu_\nu=\frac{1}{2}\kappa_5^2\left({}^{(4)}\!T^\mu_\nu
-\frac{1}{3}{}^{(4)}\!T\delta^\mu_\nu\right)
=\mp\frac{\sqrt{2}\alpha}{6V}\delta^\mu_\nu.
\end{equation}
Further, the boundary condition on $V$ at the brane is expressed as
\begin{equation}
\nabla_\perp V=\pm \sqrt{2}\alpha.
\end{equation}
Hence we obtain
\begin{equation}
{}^{(4)}\!R(g+\langle\delta_2g\rangle)= {}^{(4)}\!R(g)
-2\kappa_5^2\langle T^{\rm GW}{}^5_5\rangle 
+ \frac{\dot B^2}{a^2}\left(\frac{\langle\delta_2 V\rangle}{V}
-\frac{\langle\dot \delta_2 V\rangle}{\dot V}\right),
\end{equation}
where from the Appendix $\langle T^{\rm GW}{}^5_5\rangle$ is given by
\begin{equation}
\kappa_5^2\langle T^{\rm GW}{}^5_5\rangle =
-\frac{3}{2a^2}\left(\dot H_T^2-k^2H_T^2\right)
+G^5_5 H_T^2
+\frac{2}{a^2}(\dot A+\dot B) H_T \dot H_T
-\frac{2}{b^2}H_T H_T'',
\end{equation}
with $b=e^{B(t)}C^2$ and 
\begin{equation}
G^5_5=\frac{1}{4b^2}\frac{(V')^2}{V^2}+\frac{1}{4a^2}\frac{\dot 
V^2}{V^2}-\frac{\alpha^2}{6V^2}
=\frac{\alpha^2}{3C^6}t^{-2\delta}+\frac{\delta^2}{4a_0^2C}t^{\delta-3}.
\end{equation}

By putting the asymptotic estimates for $H_T$ into this expression, we find 
that in the $(\downarrow)$ background case, the $\dot H_T^2/a^2$ term 
dominates and decays as 
$\tau^{-(3+\delta)/(3-\delta)}=\tau^{-1+2|\delta|/(3+|\delta|)}$. Since the 
background value of ${}^{(4)}\!R(g)$ decreases in proportion to 
$1/\tau^2\sim t^{\delta-3}$, $\langle T^{\rm GW}{}^5_5\rangle$ decreases 
more slowly than ${}^{(4)}\!R(g)$. Thus in this case 
$\langle T^{\rm GW}{}^5_5\rangle$ dominates the background value for 
${}^{(4)}\!R(g)$ in the late stage, and in order for the FRW nature of the 
brane to be preserved, $\delta_2 V/V$ must become much larger than unity. 
This implies that the backreaction of the tensor perturbation significantly 
modifies the evolutionary behavior of the 4-dimensional universe on the 
brane.


\section{Conclusion and Discussion}

In the present paper we have studied the evolution of gravitational wave 
perturbations in the dynamical FRW brane-world models of the five-dimensional 
Ho\v{r}ava-Witten theory. As the background spacetime, we have used two 
cosmological solutions, \textit{i.e.} the $(\uparrow)$ and 
$(\downarrow)$ solutions, in which the branes represent the spatially flat 
expanding FRW universes. The most important feature of these solutions was 
the fact that we can solve the evolution equation for the tensor 
perturbation with help of the method of separation of variables in spite 
of the dynamical nature of the brane. Thus we were able to study the 
spatial and the temporal behavior of the tensor perturbation separately.

Since the model is compact in the fifth dimension, we obtained a discrete 
spectrum for the separation constant which can be interpreted as the 
graviton mass, and wavefunctions for the massless modes and for the 
massive modes were decoupled as in the case of the static brane solutions. 
However, we found that the spatial behavior of the wavefunctions for the 
massive modes is different from that in the Randall - Sundrum model 
\cite{RS2}: in our case they have maximums at the branes, but in RS model 
they have minimums at the branes when expressed in terms of the variable 
$H_T$ adopted in the present paper. This suggests that the coupling 
between the massless modes and the massive modes on the branes may become 
important when we consider non-linear corrections in the models considered 
in the present paper.

Although the spatial behavior of perturbations for the two solutions
was exactly the same, their temporal behavior was quite
different. Namely, we have found that in the $(\uparrow)$ background
the temporal behavior of massive modes approach that of massless modes
in a late time, while in the $(\downarrow)$ background the massive
modes decay more rapidly than the massless modes. We can understand
this difference as being caused by the difference in the behavior of $
\exp[A(t)-B(t)] $, i.e., the difference between the expansion rate of
the four-dimensional FRW brane and that of the orbifold space. Roughly
speaking, waves become massless when $\lambda\, m_{KK}\ll1$, while
they become massive when $\lambda\, m_{KK}\gg1$, where $\lambda \equiv
a_0 e^{A(t)}\sqrt{C(y)} k^{-1}$ is the reduced proper
wavelength. Therefore, in the $(\uparrow)$ background, since $
\exp[A(t)-B(t)] $ is a decreasing function and $ \lambda\, m_{KK}
\longrightarrow 0$, every mode becomes effectively massless in the
late time. On the other hand, in the case of the $(\downarrow)$
background, $\exp[ A(t)-B(t)] $ is an increasing function, and the
modes becomes more and more massive with time. Then, as the WKB
approximation shows, they suffer from an extra damping in proportion
to $1/(\lambda m_{KK})$.

This result shows that in both models the tensor perturbation is
dominated by massless modes. In models such as the Randall-Sundrum
model in which the bulk geometry is determined by a cosmological
constant, massless modes of the tensor perturbation are expected to
have no important effect on the bulk geometry. In contrast, in the
$(\uparrow)$ solution of the Ho\v{a}va-Witten theory, the energy
density of the bulk spacetime decreases in time. Hence, the
backreaction of the energy density of the tensor perturbation may
become important. In fact, we have shown that in the second-order
perturbation framework, the contribution of the tensor perturbation
supersedes the original background energy density determining the bulk
geometry in the $(\uparrow)$ case. Hence this background solution is
unstable against non-linear corrections. Although this result was
obtained for a special brane motion obtained under the assumption that
the brane contains no matter apart from the $\phi$ field, it also
holds for a more realistic brane which contains ordinary matter. It is
because the essential feature of the temporal behavior of the tensor
perturbation does not depend on the boundary condition at the
brane. 

We have also examined the 2nd-order backreaction of the tensor 
perturbation on the intrinsic geometry of the branes, and have found that 
the backreaction effect significantly alters the time evolution of the 
brane geometry for the $(\downarrow)$ solution, although the backreaction 
on the bulk geometry is negligible for this solution. This result is 
consistent with the naive expectation that the massive modes of the tensor 
perturbation behave as dark matter. 

These results suggest that the stability against the backreaction can
be used as a criterion to physically acceptable brane-world models and
to discuss cosmological implications of models. Thus it will be
interesting to analyze the non-linear stability of the Randall-Sundrum
models as well as of more realistic solutions in the Ho\v{r}ava-Witten
theory.

\acknowledgements{}

O.S would like to thank A. Ishibashi for reading the manuscript and 
discussion. H.K was supported by by the Grant-In-Aid for the Scientific 
Research (C2) of the Ministry of Education, Science, Sports and Culture in 
Japan (11640273).

\appendix
\section*{}

In this appendix, we give the expression for the second order part
with the respect to the perturbations of the $(m+n)$-dimensional
Einstein-Hilbert action 
\begin{equation}
S = \frac{1}{2 \kappa^2} \int d^{m+n} x \sqrt{-g}(R-2\Lambda), \label{action2}
\end{equation}
and its energy-momentum tensor, where $\kappa$ and $R$ are the
$(m+n)$-dimensional gravitational constant and Ricci scalar,
respectively, and $\Lambda$ is the cosmological constant.

By decomposing the metric $g_{\mu\nu}$ into the background ${\bar
g}_{\mu\nu}$ and the perturbation $h_{\mu\nu}$ as
\begin{equation}
g_{\mu\nu} = {\bar g}_{\mu\nu}+h_{\mu\nu}.  \label{decomp}
\end{equation}
and substituting it into Eq.(\ref{action2}), we obtain the following
expression for the second order part with the respect to the
perturbation $h_{\mu\nu}$ of the $(m+n)$-dimensional Einstein-Hilbert
action:
\begin{eqnarray}
S_2 &=& \frac{1}{2\kappa^2}\int d^{m+n} x \sqrt{-{\bar g}}
\left[\frac{1}{4}\left(h^{\mu\nu}{}_{;\rho}(2h^{\rho}{}_{\mu;\nu}
-h_{\mu\nu}{}^{;\rho})+h_{;\mu}(h^{;\mu}-2h^{\mu\nu}{}_{;\nu})\right)\right.
\nonumber\\ 
& & \left. +\frac{1}{8}(h^2-2h^{\mu\nu}h_{\mu\nu}){\bar R}
+\frac{1}{2}(2h^{\mu\rho}h_{\rho}{}^{\nu}-hh^{\mu\nu}){\bar R}_{\mu\nu} 
-\frac{1}{4}\Lambda(h^2-2h^{\mu\nu}h_{\mu\nu})\right].
\end{eqnarray}
By taking the variation of this action with respect to the background
metric ${\bar g}_{\mu\nu}$, 
\begin{equation}
\delta S_2 = \int d^{m+n} x \sqrt{-{\bar g}}\frac{1}{2}T^{\mu\nu}\delta {\bar g}_{\mu\nu} ,
\end{equation}
we obtain the following energy-momentum tensor $T_{\mu\nu}$: 
\begin{eqnarray} 
4\kappa^2T_{\mu\nu} &=& -g_{\mu\nu}\left[\frac{1}{2}h^{\rho\sigma}{}_{;\lambda}h_{\rho\sigma}{}^{;\lambda}-\frac{1}{2}h_{;\rho}h^{;\rho}-h_{;\rho\sigma}h^{\rho\sigma}+h^{\rho\sigma}{}_{;\sigma}h_{\rho}{}^{\lambda}{}_{;\lambda}+2h^{\rho\sigma}{}_{;\sigma\lambda}h_{\rho}{}^{\lambda} \right. \nonumber \\
 & & \left. +\frac{1}{2}\square(h^2-2h^{\rho\sigma}h_{\rho\sigma})+(h^{\rho\lambda}h_{\lambda}{}^{\sigma}-hh^{\rho\sigma})_{;\rho\sigma}+R_{\sigma}{}^{\lambda}{}_{\rho\alpha}h^{\sigma\alpha}h^{\rho}{}_{\lambda}-(h^{\rho\lambda}h_{\lambda}{}^{\sigma}-hh^{\rho\sigma})R_{\rho\sigma}\right] \nonumber \\
 & & -\frac{1}{2}(h^2-2h^{\rho\sigma}h_{\rho\sigma})\left(R_{\mu\nu}-\frac{1}{2}g_{\mu\nu}R+g_{\mu\nu}\Lambda\right) \nonumber \\
 & & +h^{\rho\sigma}{}_{;\mu}h_{\rho\sigma;\nu}-2h_{\nu}{}^{\rho}\square h_{\rho\mu}-2h^{\rho\sigma}{}_{;\mu\rho}h_{\nu\sigma}-2h^{\rho\sigma}{}_{;\mu}h_{\nu\sigma;\rho}+2h^{\rho}{}_{\mu;\nu\sigma}h^{\sigma}{}_{\rho}+2h^{\rho}{}_{\mu;\nu}h^{\sigma}{}_{\rho;\sigma} \nonumber \\
 & & -h_{;\mu}h_{;\nu}+h_{\mu\nu}\square h+2h_{;\mu}h^{\rho}{}_{\nu;\rho}-h^{;\rho}h_{\mu\nu;\rho}-2h^{;\rho}{}_{;\nu}h_{\rho\mu}-2h_{\mu}{}^{\rho}{}_{;\rho}h_{\nu}{}^{\sigma}{}_{;\sigma}+2h^{\rho\sigma}{}_{;\sigma}h_{\mu\nu;\rho} \nonumber \\
 & & +4h^{\rho\sigma}{}_{;\sigma\mu}h_{\rho\nu} \nonumber \\
 & & +6R_{\mu}{}^{\rho}{}_{\sigma\lambda}h_{\nu}{}^{\lambda}h^{\sigma}{}_{\rho}+2(h^{\rho}{}_{\mu}h^{\sigma}{}_{\nu}-h_{\mu\nu}h^{\rho\sigma})_{;\rho\sigma}-(R-2\Lambda)(hh_{\mu\nu}-2h_{\mu\rho}h^{\rho}{}_{\nu})  \nonumber \\
 & & +\frac{1}{2}(h^2-2h^{\rho\sigma}h_{\rho\sigma})_{;\mu\nu}-4R_{\mu\rho}(h_{\nu}{}^{\sigma}h_{\sigma}{}^{\rho}-hh^{\rho}{}_{\nu})-2R_{\rho\sigma}(h^{\rho}{}_{\nu}h_{\mu}{}^{\sigma}-h^{\rho\sigma}h_{\mu\nu}) \nonumber \\
 & & -\square(h_{\mu}{}^{\rho}h_{\rho\nu}-hh_{\mu\nu})+2(h^{\rho\lambda}h_{\lambda\nu}-hh^{\rho}{}_{\nu})_{;\mu\rho} ~~~~, \label{em}
\end{eqnarray}
where $\square$ is the $(m+n)$-dimensional d'Alembertian.
 
Under the notation adopted in \cite{kodama}, the unperturbed
background geometry in brane-world models is expressed as
\begin{equation}
d{\bar s}^2={\bar g}_{\mu\nu}dz^{\mu} dz^{\nu}=g_{ab}(y)dy^a dy^b + r^2(y)d\sigma_n^2,
\end{equation}
where the metric
\begin{equation}
d\sigma_n^2=\gamma_{ij}(x)dx^i dx^j
\end{equation}
is that of the $n$-dimensional space with a constant sectional
curvature $K$. The tensor mode of the metric perturbation is expanded as
\begin{equation}
h_{ab}=0,\quad h_{ai}=0,\quad h_{ij}=2r^2 H_{Tij}.
\end{equation}
For this tensor perturbation, the spatial average of the energy-momentum
tensor (\ref{em}) in the spatially flat ($K=0$) case is given by
\begin{eqnarray}
& \kappa^2\langle T_{ab}^{GW} \rangle = 
& g_{ab}\left[\frac{3}{2}\left\{(DH_T)^2+\frac{k^2}{r^2}H_T^2\right\}
+\frac{2}{r}H_T Dr\cdot DH_T + \Lambda H_T^2\right] \nonumber\\
&& -D_aH_T D_bH_T-2H_TD_aD_bH_T-\frac{4}{r}H_TD_arD_bH_T+G_{ab}H_T^2,\\
& \kappa^2\langle T_{ai}^{GW} \rangle =&0,\\
& \kappa^2\langle T_{ij}^{GW} \rangle =
& r^2\delta_{ij}\left[\frac{3}{2}\left\{(DH_T)^2+\frac{k^2}{r^2}H_T^2\right\}
+\left(\Lambda+\frac{1}{n}G^k_k\right)H_T^2\right]\nonumber\\
&& -\left[2\left\{r^2(DH_T)^2+k^2H_T^2\right\}+4r^2\left(\frac{1}{n}G^k_k+\Lambda\right)H_T^2\right]C_{ij}
+k_ik_j H_T^2,
\end{eqnarray}
where we have normalized as $\langle
\mathbb{T}$${}_{ij}\mathbb{T}$${}^{ij}\rangle = 1$ and $\langle
\mathbb{T}$${}_{ik}\mathbb{T}$${}^k{}_j\rangle = C_{ij}$.


\end{document}